\documentstyle[prl,aps,twocolumn,epsfig,floats]{revtex}

\begin{document}
\draft
\twocolumn[\hsize\textwidth\columnwidth\hsize\csname %
@twocolumnfalse\endcsname

\preprint{}

\title{An Alternative Explanation on the Two Relaxation Rates in Cuprate Superconductors}
\author{Nie Luo}
\address{Department of Physics and Astronomy, Northwestern University, Evanston, IL 60208}
\date{\today}
\maketitle
\begin{abstract} 

Transport properties of high transition temperature (high-$T_c$) superconductors have been shown to have two distinct relaxation rates. We argue that this apparent inconsistence can be resolved with an effective carrier density $n$ linear in temperature $T$. Experimental evidences both for and against this explanation are analyzed and we conclude that this offers a simple yet promising scenario. Above the pseudogap temperature $T^*$ the normal state of high-$T_c$ cuprates is close to a Fermi liquid. It however assumes non-Fermi-liquid behaviors below $T^*$. 
\end{abstract}
]
\narrowtext

The peculiar normal-state transport properties of high-$T_c$ superconductors are rather controversial and not well understood. The most striking of these is the observation of two relaxation rates \cite{anderson}. The resistivity $\rho$ is linear in temperature $T$, which implies a transport (longitudinal \cite{romero-cyclo}) relaxation rate $1/\tau_{tr}\propto T$ from $\rho=m_c/ne_c^2\tau_{tr}$ assuming that $e_c$, $m_c$, $n$, respectively the charge, mass and density of carriers are $T$-independent. In contrast, the Hall (transverse) relaxation rate $1/\tau_H$ from the cotangent of Hall angle $\cot\theta_H=m_cc/e_cH\tau_H$ is quadratic in $T$.  These two distinct $T$ dependences have various explanations. Probably the best known is due to Anderson, based on the spin-charge separation of a Luttinger liquid \cite{anderson}. Other interesting scenarios include but are not limited to the near-antiferromagnetic Fermi liquid (NAFL) theory of D. Pines {\it et al.} \cite{pines} and marginal Fermi liquid (MFL) theory by Varma {\it et al.} \cite{varma}. 

We are here to point out that the two distinct rates might be superficial if the effective carrier density (concentration) $n$ actually depends on $T$, or $n=n(T)$. Especially if the $n(T)\propto T$, then we are left with perhaps only one rate, for both longitudinal and transverse processes.  This is easy to see because $1/\tau_{tr}=n(T)e_c^2\rho/m_c$ (in Gaussian units) and as long as $n(T)\propto T$ and $\rho \propto T$ we then get $1/\tau_{tr} \propto T^2$.  It has therefore the same $T$ dependence as $1/\tau_H$ from the Hall effect. 

Similar explanations have been proposed before \cite{hirsch,ykubo,xing}. Alexandrov and Mott also suggest $n(T) \propto T$ although their carriers are bosons (bipolarons) \cite{alexandrov,alexandrov1}. This scheme of variable $n$ however does not come without difficulties. Optical conductivity \cite{romero-IR,quijada,IR-fit} seems to suggest that spectral weight of the Drude part is independent of $T$,implying a constant $n$. Nevertheless, we will show later that such an evidence is oversimplified. 

Although the discrepancy between $\rho\propto T$ and $\cot \theta _H \propto T^2$ is widely treated as abnormal, similar behaviors happen elsewhere, typically in some semimetals. Bismuth displays a similar discrepancy from 77 K to 300 K as shown in Figs. 1(a) and 1(b), although with a $\cot(\theta _H)\propto T^{2.5}$ instead \cite{michenaud}. This behavior has long been understood from a single rate $1/\tau \propto T^{2.5}$ and a variable $n(T) \propto T^{1.5}$ [see Fig. 1(c)], which in combination give a $\rho \propto T$. It is very {\it unlikely} that a Luttinger liquid theory is relevant to a three-dimensional conductor, to which Bi belongs.  Then taking such discrepancy as the evidence for spin-charge separation is {\it not} persuasive. 

\begin{figure} [t]
\begin{center}
\leavevmode
\epsfxsize=3.3in 
\epsfysize=2.9in 
\epsffile{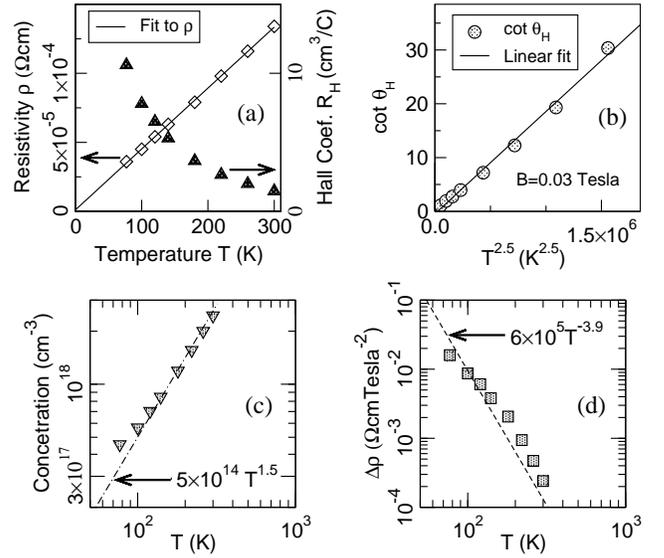}
\end{center}
\caption{(a) Resistivity $\rho$($\rho_{33}$) and Hall coefficient $R_H$($R_{231}$) versus $T$ in Bi after Ref. [14]. (b)  $\cot\theta_H$ versus $T^{2.5}$ calculated from $\rho_{33}$ and $R_H$($R_{231}$). (c) Concentration for both electron and hole versus $T$. (d) Magnetoresistance $\Delta\rho$($A_{33}$) versus $T$.}
\label{fig1}      
\end{figure}

With a much lower $n$ ($\sim 10^{17}$ cm$^{-3}$), the semimetal Bi still has a lower resistivity than most cuprates in their normal states, which are often treated as metals. This raises the question of whether doped cuprate is a metal,  a semimetal, or even a semiconductor as suggested by Alexandrov and Mott \cite{alexandrov1,alexandrov-dual}. We believe that it has the characters of all the three. It is like metal or semimetal as required by a non-zero $n$ when approaching $0\,{\rm K}$. On the other hand, significant electrostatic field effects \cite{electrostatic} on the transport, the divergent $0\,{\rm K}$ resistance under a strong magnetic field \cite{lowTstrongH} and the likely variable $n$ from the Hall effect \cite{HallwithT} put it near semiconductors and semimetals, quite far from a simple metal \cite{simplemetal} where $n$ is taken as fixed. 

Let us go back to other evidences supporting the $n(T) \propto T$ argument. The first and foremost is actually the Hall coefficient $R_H$ itself. Within the temperature range where $\cot\theta_H \propto T^2$,  $R_H$ can be nicely fitted by $1/(a_0+a_1T)$ in optimally doped (OD) $p$-type cuprates \cite{HallwithT}. Note that $R_H=1/ne_cc$ or equivalently $1/e_ccR_H$ measures $n$ in a simple parabolic band. However a naive explanation of $n=1/e_ccR_H$ seems not prudent because $R_H$ may reach 0 in some cuprates. This is perhaps why the $T$-dependent $R_H$ is not widely accepted as the evidence for $n(T) \propto T$. However $R_H=0$ can be understood with a two-carrier model as a result of compensation \cite{luo,eagles,xing-2band,hirsch1}. And as long as one carrier dominates the transport, $R_H$ is still $\propto 1/n $ although $e_ccR_H=1/n$ no longer holds. To see this, suppose that the densities of two carriers are given respectively by $n_e=n_{e0}+n_{e1}T$ and $n_h=n_{h0}+n_{h1}T$, giving rise to a Hall coefficient
\begin{eqnarray}
R_H &=& \frac{n_h-n_eb^2}{e_cc(n_h+n_eb)^2} \nonumber \\ 
 &=& \frac{(n_{h0}-n_{e0}b^2)+(n_{h1}-n_{e1}b^2)T}{e_cc\left[ (n_{h0}+n_{e0}b)+(n_{h1}+n_{e1}b)T \right] ^2},  \nonumber
\end{eqnarray}
where $b=\mu_e/\mu_h$ is the mobility ratio. If $n_{e1}T \gg n_{e0}$, $n_{h1}T \gg n_{h0}$, and $n_{h1}-n_{e1}b^2 \neq 0$, $R_H$ is roughly $\propto 1/T$. Thus the Hall effect strongly suggests $n(T) \propto T$ \cite{caution} in the normal state of OD $p$-type cuprates.  

$R_H$ is arguably the most widely used and quite accurate method to determine $n$ in metals and semiconductors when compensation is not severe. It is not persuasive to treat one half of the Hall effect, say $\cot\theta_H \propto T^2$ as exact, while disregarding the physical meaning of the other half, namely $R_H \propto 1/T$, for $\cot \theta_H = m_cc/e_cH\tau_H$ itself holds strictly only for a simple parabolic band, just as $R_H=1/ne_cc$. Actually putting these two facts at equal footing makes the physics compact and concise: there roughly exists a relaxation rate $1/\tau \propto T^2$, governing both longitudinal and transverse processes. $\rho=m_c/n(T)e_c^2\tau_{tr} \propto T$ is the result of $1/\tau_{tr} \approx 1/\tau \propto T^2$, and $n(T) \propto T$, while $\cot\theta_H = m_cc/e_cH\tau_H$ has no $n(T)$ in it thus $\cot\theta_H \propto 1/\tau_H \approx 1/\tau \propto T^2$. (We assume that $m_c$ is relatively $T$-independent and this issue will be addressed later). With a non-fixed $n$, we can proceed on to the apparent violation of Kohler's rule \cite{kohlersRule}. 

The Kohler's rule in its ordinary form states that the relative magnetoresistance $\Delta\rho/\rho_0$ in a magnetic field $H$, can be represented in the form 
\cite{zimanKohler}
\begin{equation}
\Delta\rho/\rho_0=F(H/\rho_0)\label{kohlerEq1},
\end{equation}
where $\rho_0$ is the resistivity at $H=0$ and $F$ is a function given by the metal and its sample geometry only.  
 
In cuprates, approximately $\Delta\rho/\rho_0\propto H^2T^{-4}$ \cite{kohlersRule} but $H/\rho_0\propto HT^{-1}$. Because $\Delta\rho/\rho_0$ (or $H^2T^{-4}$) is not a function of $H/\rho_0$ (or $HT^{-1}$) only, Kohler's rule seems to be violated. 

The argument above is however not valid if carrier density $n$ is not fixed. A scrutiny on its derivation \cite{zimanKohler} shows that a more general form of Kohler's rule is 
\begin{equation}
\Delta\rho/\rho_0=F(H\tau)\label{kohlerEq2},
\end{equation} 
where $\tau$ is the single relaxation time assumed for that conductor. For simple metals, $n$ is roughly a constant, so that $\tau$ has the same $T$ dependence as $\rho_0$ and thus there is no problem if one writes $\Delta\rho/\rho_0=F(H/\rho_0)$ instead of $F(H\tau)$. However for cuprates, $n$ likely depends on $T$, so we should use Eq. (\ref{kohlerEq2}) instead of Eq. (\ref{kohlerEq1}) in order to see if Kohler's rule survives. Because $H\tau \propto HT^{-2}$ where $1/\tau$ is the single rate assumed, we have $\Delta\rho/\rho_0 \propto H^2T^{-4} = (HT^{-2})^2$ and thus $\Delta\rho/\rho_0 \propto (H\tau)^2 = F(H\tau)$ so that Kohler's rule still holds. [Similarly in Bi, $\Delta\rho/\rho_0 \propto H^2T^{-5} = (HT^{-2.5})^2$ inferred from Fig. 1(d) and so we get $\Delta\rho/\rho_0 \propto (H\tau)^2 = F(H\tau)$ because $1/\tau \propto T^{2.5}$ here in Bi. Kohler's rule is not violated either.]

\begin{figure} [t] 
\begin{center}
\leavevmode
\epsfxsize=3.4in 
\epsfysize=3.3in 
\epsffile{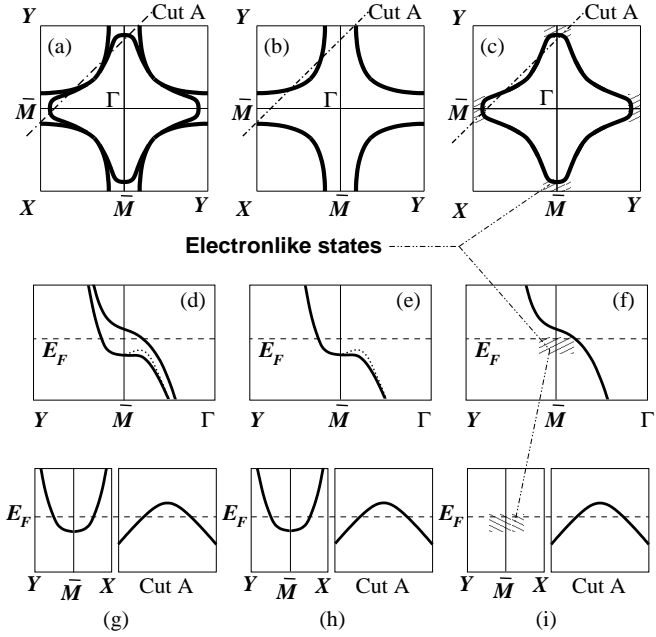}
\end{center}
\caption{Fermi surface and band structure scenarios of Bi-2212 from ARPES. (a) FS due to split from two CuO$_2$ planes. (b) Holelike FS shared by all cuprates. (c) Holelike FS with electronlike states shown as the shaded area. (d) (e) (f) Band structures near FS along $Y$-${\bar M}$-$\Gamma$ for cases shown in (a), (b) and (c) after Ref. [26] Dotted lines along ${\bar M}$-$\Gamma$ in (d) and (e) indicate hypothesized band dispersion which stabilizes local minimum at ${\bar M}$. (g) The lower band of (d) unfolded along $Y$-${\bar M}$-$X$ on the left panel. The band is electronlike near ${\bar M}$ along $Y$-${\bar M}$-$X$. (h) Band of (e) unfolded along $Y$-${\bar M}$-$X$ on the left panel, electronlike. (i) Electronlike states near ${\bar M}$ on the left panel. The right panels of (g), (h) and (i) show holelike band dispersion near $(\pi/2, \pi/2)$ along Cut A, indicated by dash-dotted lines in (a), (b) and (c). The possible complication in holelike bands from the bilayer splitting near $(\pi/2, \pi/2)$ as in (a) is neglected. $E_F$ is the Fermi level in all cases.}
\label{fig2}      
\end{figure}  

The perseverance of Kohler's rule barely strengthens the idea of a single rate $1/\tau \propto T^2$. Thus our previous explanation of the Hall effect seems to be in the right direction. Angle-resolved photoemission spectroscopy (ARPES) studies also suggest the applicability of semimetal band structure to cuprates. Fig. 2 shows some likely scenarios how this might be materialized in Bi-2212 with the Fermi surface (FS) mapping following Ref. \cite{arpes}. The discussion can be generalized to other $p$-type superconductors.  

Having seen some evidences for this explanation, we now discuss experiments at odds with it. The most serious one, in our opinion, is from the optical conductivity $\sigma(\omega)$. To account for the non-Drude behavior of $\sigma(\omega)$ in cuprates, a classical two-component model \cite{2componentIR} is used in quite some studies, 
\begin{equation}
\sigma(\omega,T)=\sigma_{\rm D}(\omega,T) + \sigma_{\rm MIR}(\omega,T)\label{EQIR1}, 
\end{equation} 
where $\sigma_{\rm D}(\omega,T)=[n(T)e_c^2/m_c]\{(1/\tau)/[\omega^2+(1/\tau)^2]\}$ is the $T$-dependent Drude part while $\sigma_{\rm MIR}(\omega,T)$ is the {\it nearly} $T$-independent midinfrared (MIR) part. The spectral weight of Drude part $\int_0^\infty \sigma_{\rm D}(\omega,T) d\omega$, is found to be independent of $T$ [the shape of $\sigma_{\rm D}(\omega, T)$ is $T$-dependent though], suggesting a constant $n$ if $m_c$ is taken independent of $T$. Nonetheless the typical use of such a model is oversimplified. The problem is: the MIR is treated as $T$-independent and modeled as classical Lorentz oscillators.  

\begin{figure} [t]
\begin{center}
\leavevmode
\epsfxsize=3.2in 
\epsfysize=2.9in 
\epsffile{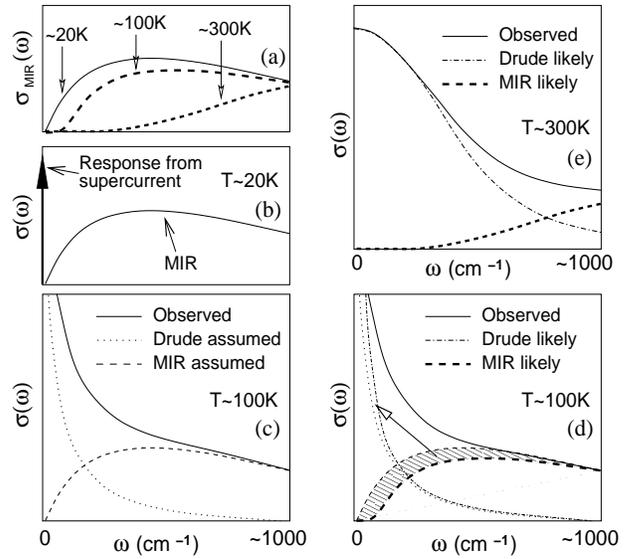}
\end{center}
\caption{Spectral weight transfer between the Drude and MIR part of optical conductivity $\sigma(\omega,T)$ in cuprates. (a) Theoretical variation of $\sigma_{\rm MIR}(\omega,T)$ with $T$ in quantum model. (b) The $\sigma(\omega)$ as found at $20\,{\rm K}$. (c) $\sigma(\omega)$ at $\sim 100\,{\rm K}$. The dashed line is the $\sigma_{\rm MIR}(\omega,T)$ assumed in the typical two-component analysis, which is the same as $\sigma_{\rm MIR}(\omega)$ at $20\,{\rm K}$. The dotted line is the calculated $\sigma_{\rm D}(\omega)$ by using $\sigma(\omega, 100\,{\rm K})- \sigma_{\rm MIR}(\omega, 20\,{\rm K})$. (a), (b) and (c) share the same horizontal axis. (d) $\sigma(\omega)$ at $\sim 100\,{\rm K}$. The thick dashed line is the actual $\sigma_{\rm MIR}(\omega)$ at $100\,{\rm K}$ as shown in (a) as the long dashed line. The dash-dotted line is the actual $\sigma_{\rm D}(\omega)$ at $100\,{\rm K}$ by using $\sigma(\omega, 100\,{\rm K})- \sigma_{\rm MIR}(\omega, 100\,{\rm K})$. The thin dashed and dotted lines are the same as those in (c) for comparison. Obviously the Drude weight obtained in (d) is larger than that in (c), which means an higher carrier density at $100\,{\rm K}$ compared with $20\,{\rm K}$. The shaded area is the spectral weight transfered away from MIR as indicated by the arrow when $T$ reaches $100\,{\rm K}$ from $20\,{\rm K}$. (e) $\sigma(\omega)$ at $\sim 300\,{\rm K}$. The spectral weight of MIR is further transfered to the Drude part.}
\label{fig3}      
\end{figure}

The $T$-dependence of MIR is tied to its nature. MIR is often understood as an interband electronic transition \cite{interbandexcitation,interbandexcitation1}. This interband transition, especially at its low $\omega$ part, is however not from excitation over the charge-transfer (CT) gap. Rather it perhaps comes from doping-related states and spectral weight transfers inside the CT gap as suggested by various doping-dependent photoemission, inverse photoemission, X-ray absorption, electron-energy-loss spectroscopy and IR reflectance studies \cite{dopingstates,interbandexcitation1}. Without lingering over the exact origin of these MIR-related states, we have two observations from experiments. First, MIR states are very close to the Fermi level, manifested by low energy MIR tail approaching $\omega =0$ \cite{MIRlow}. Such states might relate to the flat band, van Hove singularity and electronlike states [see Figs. 2(c) and 2(f)] revealed in ARPES. Second, MIR states are localized at low $T$ as seen in $\sigma(\omega)$ at 20 K or lower \cite{romero-IR,quijada} as shown in Figs. 3(b), because otherwise it would behave Drudelike and there would not be an MIR at all. Then the interband transitions of MIR are likely from shallow localized states to extented ones.

With these two properties of MIR states, $\sigma_{\rm MIR}$ should be $T$-dependent in its low $\omega$ part (actually in far infrared, FIR), because higher $T$ should progressively set more carriers free, which corresponds to a spectral weight transfer from MIR (or FIR) to the Drude part as indicated by the arrow in Fig. 3(d). In other words, the classical model of Lorentz oscillator with a $T$-independent MIR is {\it not} correct considering the quantum nature of interband excitations. With increasing $T$, there should be a reduction in the spectral weight of MIR as shown in Fig. 3(a), and this reduction in MIR should then be made up by an increase in the Drude part [Fig. 3(d)] according to the sum rule ({\it i.e.}, a spectral weight transfer).  Such weight reduction \cite{weightreduction} and weight transfer \cite{weighttransfer} have been understood in optical studies of narrow-gap semiconductors like HgTe, Hg$_{1-x}$Cd$_x$Te, and semimetals such as Bi. Unlike valence-to-conduction-band type of transition there ($\Gamma_8$ $\to$ $\Gamma_8$ in HgTe, Hg$_{1-x}$Cd$_x$Te and possibly crossing the $L$-point gap in Bi), the MIR in cuprates are likely from localized to extended bands, but a similar $T$-dependence should still hold. However being overlapped $\sigma_{\rm D}$ and $\sigma_{\rm MIR}$ in cuprates have no clear-cut method to separate. (Overlap of the two parts in Hg$_{1-x}$Cd$_x$Te and Bi is not as severe.)  So what is often used in analysis, is assuming that the $\sigma_{\rm MIR}$ at 20 K or lower, written indiscriminately as $\sigma_{\rm MIR}(\omega,20\,{\rm K})$ for simplicity, be the same for all other $T$ \cite{romero-IR,quijada}. Such a procedure without little doubt would make a constant Drude weight simply because of the conductivity sum rule.

Rewrite Eq. (\ref{EQIR1}) as
\begin{equation}
\sigma_{\rm D}(\omega,T)= \sigma(\omega,T) - \sigma_{\rm MIR}(\omega,T)\label{EQIR2}.
\end{equation} 
Integrate both sides over $\omega$ from 0 to $\infty$, and by using sum rule for $\sigma_{\rm D}(\omega,T)$, we get 
\begin{equation}
\frac{\pi n(T)e_c^2}{2m_c}= \int_0^{\infty}\sigma(\omega,T)d\omega - \int_0^{\infty}\sigma_{\rm MIR}(\omega,T) d\omega\label{EQIR3},
\end{equation}
where the total weight $\int_0^{\infty}\sigma(\omega,T)d\omega$ of the two parts basically measures the number of states available within the CT gap \cite{cautionIR}. It is directly integrated from experimental $\sigma(\omega,T)$ and is found independent of $T$. If a $\sigma_{\rm MIR}(\omega,T)=\sigma_{\rm MIR}(\omega,20\,{\rm K})$ is assumed for all $T$, as is done in a typical $T$-dependent two-component analysis \cite{romero-IR,quijada}, we are left with  
\begin{equation}
n(T) = \frac{2m_c}{\pi e_c^2} \left[ \int_0^{\infty}\sigma(\omega,T)d\omega - \int_0^{\infty}\sigma_{\rm MIR}(\omega,20\,{\rm K}) d\omega \right]\label{EQIR4}.
\end{equation}

Because $\int_0^{\infty}\sigma_{\rm MIR}(\omega,20\,{\rm K}) d\omega$ is a constant, the right hand side of Eq. (\ref{EQIR4}) is $T$-invariant, and we are left with an effective carrier density $n \propto \int_0^{\infty}\sigma(\omega,T)d\omega - \int_0^{\infty}\sigma_{\rm MIR}(\omega,20\,{\rm K}) d\omega$ which is independent of $T$ even if $n$ changes with the temperature in reality. Introduction of self-consistent iteration \cite{quijada} marginally improves the result but qualitatively it would not cure the problem. There are different approaches using straightforward least-squares fit \cite{quijada,IR-fit}. However, with a classical and oversimplified Drude-Lorentz model to start from, such fits are not likely to uncover the sizable $T$-dependence of the Drude weight.

Meanwhile the high $\omega$ part of MIR is less affected by $T$, as easily understood from quantum statistics and because the MIR is several times larger than Drude part, a spectral transfer on the order of one Drude weight does not contradict to the convention of nearly $T$-independent MIR. The transfer only occurs at low $\omega$ where the overlap with Drude peak makes its detection hard. 

The analysis above explains the origin of variable $n$ as well: as $T$ increases, previously localized carriers are now set free and we are likely left with an $n \propto T$. Actually this variable $n$ finds common point in the one-component model of $\sigma(\omega, T)$ \cite{1-component}, where renormalized relaxation rate $1/\tau^*$ and renormalized mass $m_c^*$ both depend on $T$ and $\omega$. In other words,
\begin{equation}
\sigma(\omega,T)=\frac{ne_c^2}{m_c^*(\omega,T)}\frac{1/\tau^*(\omega,T)}{\omega^2+[1/\tau^*(\omega,T)]^2}\label{1compEq1},
\end{equation} 
where $n$ is however taken as fixed. In a typical result of this model \cite{romero-1-comp}, the renormalized mass $m_c^*(\omega, T)$ at low $\omega$ say $200\,{\rm cm}^{-1}$ decreases with increasing $T$ as $m_c^*(\omega,T) = m_{c0}a_0/(a_0+a_1T)$ with $m_{c0}$ the $m_c^*$ at low $T$ while the renormalized rate $1/\tau^*(\omega, T)$ at low $\omega$ roughly increases as $T^2$. This behavior is perhaps equivalent to an alternative combination of low frequency $n(\omega, T) \propto (a_0+a_1T)/a_0$ and $1/\tau^*(\omega,T) \propto T^2$ if the mass $m_c$ is taken as fixed instead. Mathematically, by inserting $m_c^*(\omega,T) = m_{c0}a_0/(a_0+a_1T)$ in Eq. (\ref{1compEq1}), we have 
\begin{eqnarray}
\sigma(\omega,T) &=& \frac{ne_c^2}{m_{c0}a_0/(a_0+a_1T)}\frac{1/\tau^*(\omega,T)}{\omega^2+[1/\tau^*(\omega,T)]^2} \\
&=& \frac{n[(a_0+a_1T)/a_0]e_c^2}{m_{c0}}\frac{1/\tau^*(\omega,T)}{\omega^2+[1/\tau^*(\omega,T)]^2}.
\end{eqnarray} 

If we treat mass $m_{c0}$ as fixed, then $n(a_0+a_1T)/a_0$ can be taken an effective carrier density which is linear in $T$, and we arrive at the same conclusion as we made earlier. There must exist renormalization effects to some extent, but meanwhile we cannot rule out the variation of $n$ with $T$. The problem is however that the current optical technique is not able to differentiate one effect from the other because $m_c$ and $n$ are entangled together in the expression of optical conductivity.       

Electronic specific heat $C^{el}$ seems to suggest a fixed $n$ because $\gamma = C^{el}/T$ is a constant above $T_c$ \cite{elspecificheat}. Localized states, as long as not far from the Fermi surface ($\sim kT$), nevertheless contribute to $\gamma$ because these carriers may increase their energy and jump to extended states simply by thermal excitation. Classically this is interpreted as while being localized and deprived of translational degrees of freedom (DOF), they nevertheless have oscillatory DOF. Thus these states still contribute a term $\propto T$ to $C^{el}$, such that significant variation in $\gamma/T$ is not seen. There are recent reports claiming a scattering rate linear in $T$ shown by ARPES \cite{linearARPES,linearARPES-prl}. However we need to caution on the meaning of such a rate and its relation to the actual transport relaxation rate. Averaging to more than 0.2 eV at 300 K \cite{linearARPES-prl}, such rate is far from $1/\tau_{tr}$ of 300-500 cm$^{-1}$ given by current IR transport studies. On the other hand recent dynamic conductivity experiment clearly show a transport rate $1/\tau_{tr} \propto T^2$ by THz technique \cite{Thz}, which in principle is a much finer probe than ARPES in energy resolution, and more important, directly measures transport rates. The $T^2$-rate revealed in THz experiments could well be the single rate we proposed in this paper.

All these experimental evidences suggest a promising combination of $1/\tau \propto T^2$ and $n(T) \propto T$ in explaining the peculiar transport in cuprates.  Then what is the scattering mechanism behind the $T^2$ rate? It is perhaps mainly caused by electron-electron and electron-hole scatterings. A fermion-fermion scattering results in a $T^2$ rate basically because of the phase space restraints from the Pauli principle. In two-dimension (2D), some nesting effect might be prominent but it has been shown that $e$-$e$ scattering still basically follows a $T^2$ law \cite{T2}. 

Here we have however a variable $n(T)$ which might make us suspect a rate $1/\tau$ increasing faster than $T^2$. Intuitively the electrons (holes) are getting more crowded with increasing $T$. So we want to use a very simple argument based on power analysis to show that this worry is not needed.

Treating the scattering process using Fermi Golden rule, the probability per unit time that an electron in ${\bf k}$ will be scattered into another ${\bf k'}$ is given by
\begin{equation}
w({\bf k'}, {\bf k})=\frac{2\pi}{\hbar}\langle {\bf k'}|H'|{\bf k}\rangle^2\delta(\varepsilon_{{\bf k}} - \varepsilon_{{\bf k'}}),
\end{equation} 
where $|{\bf k}\rangle=e^{i{\bf k}{\bf r}}$ up to a normalization constant and $H'=e^{-\alpha r}/r$, the screened coulomb potential. In two-dimension (2D),
\begin{eqnarray}
\langle {\bf k'}|H'|{\bf k}\rangle &=& \int \frac{e^{i({\bf k} - {\bf k'}){\bf r}}e^{-\alpha r}}{r} d^2 r \\
&=& \int \frac{e^{i{\bf q}{\bf r}}e^{-\alpha r}}{r} rdrd\theta,
\end{eqnarray}
where ${\bf q}={\bf k}-{\bf k'}$ and thus $q=2k\sin(\theta/2)$ with $\theta$ the angle between ${\bf k}$ and ${\bf k'}$.

Notice that in 2D, $e^{i{\bf q}{\bf r}}$ can be expanded by Bessel functions of the first kind. Then
\begin{eqnarray}
e^{i{\bf q}{\bf r}} &=& e^{iqr\cos\theta}\\
&=& \sum_{m=-\infty}^{\infty}J_m(qr)i^m e^{im\theta}.
\end{eqnarray}

Integrate the above first over $\theta$ and $e^{im\theta}$ averages to 0 except when $m=0$. Thus 
\begin{equation}
\langle {\bf k'}|H'|{\bf k}\rangle = 2\pi\int_0^{\infty}J_0(qr)e^{-\alpha r} dr 
= \frac{2\pi}{(q^2+\alpha^2)^{\frac{1}{2}}}.
\end{equation}

When the screening is not strong which should be the case for cuprates \cite{strongcorl} we see that $\langle {\bf k'}|H'|{\bf k}\rangle^2 \approx 4\pi^2/q^2 = \pi^2/sin^2(\theta/2)k^2$, where $k$ will be taken as $k_F$ because effective scatterings only occur near the FS. To be illustrative, we only consider an isotropic case ({\it i.e.}, the FS is a circle), then the Fermi wave-vector $k_F=[2\pi n(T)]^{1/2}$ in 2D where $n(T)$ is taken as the area density of carriers. Without considering the complication from umklapp scattering, we conclude that the transition rate from ${\bf k}$ to ${\bf k'}$, $w({\bf k'}, {\bf k})$ is $\propto 1/T$ because $k^2 = k_F^2 \propto n(T)$ and $n(T) \propto T$. Now let us look at the phase space available for scattering. The perimeter of the FS is $2\pi k_F \propto [n(T)]^{1/2} \propto T^{1/2}$ and thus the number of states available for scattering to and from is proportional to the product of the perimeter $2\pi k_F$ and the thermal excitation width $\sim k_BT$ with $k_B$ the Boltzmann's constant, in other words proportional $T^{3/2}$. Applied to both initial and final states this gives us a factor of $T^3$ from the phase space restriction. Combined with a transition probability $w({\bf k'}, {\bf k}) \propto T^{-1}$ as already seen, it results in a relaxation rate $1/\tau \propto T^2$, which is exactly what we find in experiments. Complication from screening, umklapp process and so on will be shown in a separate publication but the result is essentially the same.

Similar argument applies to $e$-$h$ process where the Fermion nature of $e$ and $h$ on their own part gives the same phase restriction we mentioned before and thus the same $T$ dependence for charge transports. 

This $e$-$h$ liquid is perhaps near a Fermi liquid (FL) above the pseudogap temperature $T^*$.  Nonetheless it is not so under $T^*$ because the presence of excitonic states formed by $e$ and $h$ is significant at low $T$. These excitons are stable against recombination because of the semimetallic band structure, unlike those of fast-recombining type found in semiconductors. These bound states, as a discontinuous change from the free particle motions, manifest the loss of one-to-one correspondence from the states of free Fermi gas. This offers an explanation for various non-Fermi-liquid phenomena of cuprates. The importance of discontinuities and non-perturbative approaches in high-$T_c$ superconductivity were very well summarized by Anderson \cite{anderson1} although we do not think that spin-charge separation is the necessary step in understanding the normal state. Some 30 years ago, Mott emphasized the interacting nature of such excitonic states \cite{mott}. Kohn \cite{kohn}, Halperin and Rice \cite{rice}, pointed out the possible anomalies, like excitonic insulator, CDW, SDW, antiferromagnetic correlation, and phase separation in such a strongly interacted $e$-$h$ system, which reminds us of many of the correlation phenomena and possibly the stripe phase found in high-$T_c$ materials.

Although J\'{e}rome, Rice and Kohn \cite{kohn1}, Halperin and Rice \cite{rice} also argued that the ground state of the excitonic state is an insulator, not to mention superconductivity, their conclusion is {\it only} true for a system with equal numbers of electrons and holes. If the densities of $e$ and $h$ are not equal, the ground state should be a conductor at least. The possibility of high-$T_c$ superconductivity in systems similar to this was studied by Allender, Bray and Bardeen \cite{bardeen}, and Ginzburg \cite{ginzburg}. All these suggest a possible connection of high-$T_c$ superconductivity to the excitonic states of electron-hole liquid.


\begin{references}
\bibitem{caution} It is better to write $n(T)=a+bT$ instead of $n(T) \propto T$. We however generally neglect the difference of two such notations for the sake of simplicity.
\bibitem{cautionIR} Here $\sigma(\omega,T)$ in our notation is that from MIR and Drude only. Experimentally, this means a cutoff at $\omega$ beyond the CT gap.
\end{references}
\end{document}